\documentclass[slac_one]{revtex4}
\usepackage{graphicx}
\usepackage{fancyhdr}
\begin{document}


\title{{\small{2005 ALCPG \& ILC Workshops - Snowmass,
U.S.A.}}\\ 
\vspace{12pt}
Next Generation Multi-particle event generators for the MSSM} 

%

\author{\underline{J. Reuter}}
\affiliation{DESY Theory Group, DESY, Notkestr.~85, 22603 Hamburg, Germany}
\author{K.~Hagiwara}
\affiliation{Theory Division, KEK, Tsukuba 305--0801, Japan}
\author{W.~Kilian}
\affiliation{DESY Theory Group, DESY, Notkestr.~85, 22603 Hamburg, Germany}
\author{F.~Krauss}
\affiliation{Institute for Theoretical Physics, University of Dresden,
  01062 Dresden, Germany}
\author{T.~Ohl}
\affiliation{Dept.~of Physics and Astronomy, University of
  W\"urzburg, Am Hubland, 97074 W\"urzburg, Germany}
\author{T.~Plehn}
\affiliation{Max-Planck Institute for Physics, F\"ohringer Ring 6,
  80805 Munich, Germany}
\author{D.~Rainwater}
\affiliation{Dept.~of Physics and Astronomy, University of Rochester,
  Rochester, NY 14627, USA}
\author{S.~Schumann}
\affiliation{Institute for Theoretical Physics, University of Dresden,
  01062 Dresden, Germany}
\begin{abstract}
We present a next generation of multi-particle Monte Carlo (MC) Event
generators for LHC and ILC for the MSSM, namely the three program
packages Madgraph/MadEvent, WHiZard/O'Mega and Sherpa/Amegic++. The
interesting but difficult phenomenology of supersymmetric models at
the upcoming colliders demands a corresponding complexity and
maturity from simulation tools. This includes multi-particle
final states, reducible and irreducible backgrounds, spin
correlations, real emission of photons and gluons, etc., which are
incorporated in the programs presented here. The framework of a model
with such a huge particle content and as complicated as the MSSM makes
strenuous tests and comparison of codes inevitable. Various tests
show agreement among the three different programs; the tables of cross
sections produced in these tests may serve as a future reference for
other codes. Furthermore, first MSSM physics analyses performed with
these programs are presented here. 

\end{abstract}


\maketitle

\thispagestyle{fancy}

\section{MOTIVATION: SUSY PHENOMENOLOGY} 
\label{susypheno}

At the end of 2007, first physics data are expected to be delivered
by the Large Hadron Collider (LHC). This experiment may probe the
mechanism of electroweak (EW) symmetry breaking and possibly find
signals of new physics. There are many theoretical arguments for physics
beyond the Standard Model (SM). The most prominent among them
is low-energy supersymmetry in the form of the
minimal supersymmetric SM, the MSSM. If deviations from the SM are
seen at the LHC, it is important to distinguish which SM extension is
realized in nature. The spectrum of new particles would have to be determined
by measurements (cascade decays), the spin of new particles
assesed, and the couplings measured:
supersymmetry (SUSY) would be verified by revealing the SUSY
relations among these quantum numbers.

Furthermore, SUSY parameters would have to be determined as precisely as
possible for two reasons: simpler SUSY processes appear as backgrounds
for more complicated ones, and renormalization-group evolution
would allow us to get a handle on the GUT parameters and SUSY breaking
mechanism. The second task is part of the Supersymmetry Parameter
Analysis (SPA) project \cite{SPA}.

To match the expected experimental accuracy, which is mostly at the
per-cent level at the LHC and at the per-mille level at the ILC,
simulation tools will have to cope with the complexity of multi-particle
final states. The factorization of SUSY processes into $2\to 2$
on-shell production processes and subsequent decay of the SUSY
particles is in most cases not sufficient, since off-shell effects
can be important. Hence, off-shell intermediate states and full
gauge-invariant (sub)sets of Feynman diagrams for final states with 
six, eight and more particles have to be included. Furthermore, SM and 
MSSM backgrounds must be taken into
account, and since in general a separation of signal and
background is not guaranteed, interferences have to be accounted for. 
To identify intermediate particles in a cascade, spin information is
important. It can e.g. be accessed by the spin density matrix
formalism. 

The more traditional MC event generators like Pythia, Herwig and
SUSYGEN were designed for a different purpose, and hence do not
fulfill one or more of the abovementioned requirements. In the next
section, we present three programs which are especially designed to
fulfill these requirements.


\section{NEXT GENERATION EVENT GENERATORS FOR THE MSSM}

\subsection{Description of the tools}

In this section, we present the three multi-purpose multi-particle
event generators for the MSSM at LHC and ILC, Madgraph/MadEvent
\cite{madgraph,madevent}, WHiZard/O'Mega \cite{whizard,omega} and
Sherpa/Amegic++ \cite{sherpa,amegic}. More details about the programs and
their comparison can be found in \cite{susycomparison,higgssumm}. We
briefly describe the structure of the three codes. For matrix element
generation, where all codes use the helicity-amplitude formalism,
Sherpa first generates the topologies and then decorates them with the
particles and vertices. So the full set of Feynman diagrams is present
as a chain of subroutine calls. Madgraph is very similar. In both
programs duplicate calls are eliminated, such that identical
subamplitudes are calculated only once. O'Mega avoids all redundancies
in the matrix elements by the use of directed acyclical graphs
\cite{omega}.    

The next crucial step is phase space parameterization,
since the set of well-suited integration variables is different for
each phase space channel. The best solution is a multi-channel
adaptive integration using MC sampling. In Sherpa, the dominant
channels are selected according to the Feynman graph structure, and
the channel weights are adapted iteratively. WHiZard is quite
similar. In both programs the channel mappings are adapted iteratively
as well. MadEvent, a front-end for Madgraph, first integrates the 
single squared diagrams separately and accounts for the interferences
by correction factors afterwards. In the final step, all programs can
unweight the events, after a mapping that transforms the integrand as
closely to a constant as possible. 

The ``dressing'' of the partonic processes is done by structure
functions for the incoming partons. This makes efficient matrix
elements necessary, since integrations over two additional variables,
$x_1$ and $x_2$, are needed. For the ILC, beamstrahlung and beam energy
spread have to be taken into account to simulate a realistic
collider environment. Moreover, for polarized beams the full spin
information should be kept which is done by spin density matrices in
WHiZard. Initial state radiation (ISR) must be included, since it is
essential for the height and shape of distributions.
The parton shower of the colored particles in the final state,
hadronization and the simulation of underlying events, is performed by
an interface to Pythia in MadEvent and in WHiZard, while Sherpa
provides its own code for strong interaction effects by a systematic
merging of matrix elements with inital and final state parton showers
using the CKKW algorithm \cite{CKKW}.  

\subsection{Tests and Consistency checks of the codes}

Madgraph/MadEvent, WHiZard/O'Mega and Sherpa all read in the spectrum
and the mixing angles from a spectrum generator via the SUSY Les
Houches Accord (SLHA). For the conventions the three programs
use, cf.~\cite{susycomparison}. The MSSM is an extremely complicated model,
containing several thousands of vertices. Tests are mandatory which
guarantee their correct implementation. One 
stringent test is unitarity of scattering processes, namely
that partial-wave amplitudes for $2\to 2$ processes cannot arbitarily
grow with the energy. In many cases gauge cancellations make
the amplitudes constant (up to logarithmic corrections). This has
been checked for almost all possible $2\to 2$ and $2\to 3$ processes. 
One can also use Ward and Slavnov-Taylor identities to check the gauge
invariance of matrix elements, and even supersymmetric Ward- and
Slavnov-Tayler identities have been tested \cite{ward}. Furthermore,
an extensive comparison has been undertaken of all relevant $2\to 2$
processes necessary to test all MSSM Feynman rules.
All these tests were passed by the three programs, establishing
agreement between all the codes. The results of these tests, as well as
references for first physics results obtained with these programs
(cf. also \cite{WBF}), may be found in \cite{susycomparison}.


\section{Conclusions}

Madgraph/MadEvent, WHiZard/O'Mega and Sherpa form a new generation
of multi-particle Monte Carlo event generators for LHC and the
ILC. In the present versions, all three programs provide a full
description of the SM and the MSSM suitable for realistic physics
simulations within the collider environment, and are extensible
towards further alternative models beyond the SM. While the packages
are completely independent in their implementations of Feynman rules,
matrix element generation and phase space sampling, we found complete
agreement in a comprehensive comparison of numerical results. The
three codes are very well tested and available publicly
\cite{madgraph,madevent,whizard,omega,sherpa,amegic}. First full MSSM
physics analyses have been performed with the three programs,
including backgrounds, hadronic environments and corrections from real
emission. A major open point is the incorporation of virtual radiative
corrections into the programs with the ultimate goal of a Monte Carlo
for general processes at next-to-leading order.



\begin{acknowledgments}
T.\,O., J.\,R. and W.\,K. are supported in part by the German
Helmholtz-Gemeinschaft, Grant No.\ VH--NG--005.  T.\,O. is supported
in part by Bundesministerium f\"ur Bildung und Forschung Germany,
Grant No.{} 05HT1WWA/2.  D.R. was supported in part by the
U.S. Department of Energy under grant Nos. DOE-FG02-91ER40685,
DOE-FG02-96ER40956 and DOE-FG02-95ER40893.
\end{acknowledgments}

\end{document}